\documentclass[prb,twocolumn,amsmath,amssymb]{revtex4-1}

\usepackage{graphicx}
\usepackage{dcolumn}
\usepackage{bm}
\usepackage{color}
\usepackage{pifont}

\newcommand{\ket}[1]{ | #1 \rangle }
\newcommand{\bra}[1]{ \langle #1 | }
\newcommand{\ew}[1]{ \langle #1 \rangle }
\newcommand{\tr}{\mbox{tr}}
\newcommand{\ff}[1]{ {\boldsymbol #1} }
\newcommand{\ca}[1]{{\cal #1}}

\begin{document}
\title{One-step theory of pump-probe photoemission}
 
\author{J. Braun$^1$, R. Rausch$^2$, M. Potthoff$^2$, J. Min\'ar$^{1,3}$, H. Ebert$^1$}

\affiliation{$^1$Department Chemie, Ludwig-Maximilians-Universit\"at M\"unchen, 81377 M\"unchen, Germany \\
$^2$I.~Institut f\"ur Theoretische Physik, Universit\"at Hamburg, 20355 Hamburg, Germany \\
$^3$New Technologies - Research Center, University of West Bohemia, Univerzitni 8, 306 14 Pilsen, Czech Republic}

\begin{abstract}
A theoretical frame for pump-probe photoemission is presented. The approach is based on a general formulation
using the Keldysh formalism for the lesser Green's function to describe the real-time evolution of the electronic
degrees of freedom in the initial state after a strong pump pulse that drives the system out of equilibrium. The
final state is represented by a time-reversed low-energy electron diffraction state. Our one-step description is
related to Pendry's original formulation of the photoemission process as close as possible. The formalism allows
for a quantitative calculation of time-dependent photocurrent for simple metals where a picture of effectively
independent electrons is assumed as reliable. The theory is worked out for valence- and core-electron excitations. 
It comprises the study of different relativistic effects as a function of the pump-probe delay.
\end{abstract}

\pacs{78.47.D,78.47.J,79.60.-i}  

\maketitle

\section{Introduction}

Angle-resolved photoemission has developed over several decades into a technique of choice for determining the
electronic structure of new crystalline materials, and represents a mature tool in materials physics. \cite{pes}
Particularly, time-resolved photoemission spectroscopy (TR-PES) has been advanced on the experimental side in recent
years. To study the non-equilibrium dynamics of electronic degrees of freedom on a femto-second time scale, different
pump-probe photoemission experiments have been employed. \cite{lis05,cav07,piet08,bov10,roh11,gor11,gun11,vor12,
car12,sob12,rud12,arm12,wan12,ang13,muel13} Here, we present a general theoretical frame which can be applied to
simple metals where a treatment of the electron dynamics in a picture of essentially independent particles may be
adequate. The formalism also accounts for relativistic effects, e.g.\ it captures the simultaneous appearance of
spin-orbit coupling and magnetic exchange splitting. It can be applied to different pump-probe photoemission setups 
involving valence bands as well as core levels, in principle. To keep the complexity at a reasonable level, we focus
on a valence-pump---core-probe situation in the present study. We sketch the computation of the atomic contribution
of the initial state to the time-dependent photocurrent and point out the necessity to implement a full time-dependent 
multiple-scattering technique for other contributions to the time-dependent photocurrent.

The most successful theoretical framework available to deal with photoemission from solid surfaces is the one-step
model as originally implemented by Pendry and co-workers. \cite{Pen74,Pen76,HPT80} The main idea is to describe the
excitation process, the transport of the photoelectron to the crystal surface as well as the escape into the vacuum
\cite{BS64} as a single quantum-mechanically coherent process including all multiple-scattering events. \cite{Bra96}
Nowadays, it allows for photocurrent calculations ranging from a few eV to more than 10 keV \cite{NM11,NM12,Min13b,bra14} 
at finite temperatures and from arbitrarily ordered \cite{braun13} and disordered systems, \cite{braun10} and may
include effects of strong electron correlations in addition. \cite{braun06,pick08,barr09} However, a general
and quantitative one-step formulation of time-resolved phenomena in angle-integrated or angle-resolved photoemission 
is still missing. Only a few theoretical approaches to TR-PES have been published within the last few years. A first
description of TR-PES in terms of Keldysh Green's function techniques \cite{keldys} were published by Freericks et al.,
\cite{fre09,mor10,SKM+13} and Eckstein et al.\ \cite{eck08} followed by work from other groups. \cite{ing11,USL14}
Moreover, a first realistic description of two-photon photoemission has been worked out \cite{ueb07} as well as a
many-body formulation of core-level photoemission. \cite{fuij02}

One of the major problems, as discussed in the literature, mainly in the context of strongly correlated systems,
consists in the calculation of the lesser component of the Keldysh Green's function for a realistic system and to
avoid an equal-time approximation or similar severe simplifications. As this function has two independent time arguments,
the numerical effort can be tremendous, even for simple model systems. There is, however, another class of complications
that is relevant for the theoretical description of real materials, in particular: In order to obtain the photocurrent as
a function of the pump-probe delay, one has to calculate the lesser Green's function for a semi-infinite stack of atomic
layers and for a realistic electronic potential, typically available from band-structure formalisms like the
Korringa-Kohn-Rostoker (KKR) method. \cite{KKR} Furthermore and equally important, final-state multiple-scattering and
matrix-element effects have to be taken into account as well as the presence of the surface itself. 

For the case of {\em equilibrium} photoemission, those problems have been addressed and successfully solved in the past:
The first and most simple version of an independent-electron approximation for the photocurrent has been given by Berglund
and Spicer, \cite{BS64} namely with the so-called three-step model of photoemission where the process is divided into
three independent steps (excitation, transport, escape into the vacuum, see above). To overcome obvious deficiencies of
the three-step model, a multiple-scattering or ``dynamic'' approach has been suggested, namely first for the final state,
\cite{lie74,spa77} and later on for both, initial and final states \cite{HPT80} in order to treat self-energy corrections
on equal footing. With Pendry's one-step approach \cite{Pen76} to angle-resolved photoemission a numerically tractable
scheme was introduced which rests on the one-particle Green's function in the local-density approximation (LDA) of
band-structure theory. \cite{lda} Photoemission is described as a single coherent quantum process. Explicit effects of
strong Coulomb correlations are still disregarded. Furthermore, the use of the sudden approximation for the final state
allows to adopt an independent-particle description of the photoelectron in the framework of low-energy electron diffraction
theory. \cite{Pen74}

Our long-term goal is to provide a numerical tool which helps to analyze time-resolved pump-probe photoemission data from
real systems and which thus makes direct contact with the experiments. With the present paper, as a first step, we demonstrate
that a one-step formulation is possible in the time-dependent or non-equilibrium case, too. Using the Keldysh formalism,
\cite{keldys} the lesser Green's function provides the description of the time evolution of the electronic structure on a
femto-second time scale following a strong pump pulse. In general, this requires the solution of an integral equation involving
the $S$-matrix that corresponds to multiple (all-order) scattering at the time-dependent perturbation given by the light-matter
interaction term describing the pump. The problem can be rewritten as a Dyson equation for the non-equilibrium double-time
retarded Green's function starting from the equilibrium retarded (and advanced) Green's functions which are available from
standard KKR theory \cite{KKR} and which are homogeneous in time. However, even if one works on an LDA level and neglects
explicit Coulomb correlations in the many-body system and describes the electrons as effectively independent, this is a very
demanding task.

Within the sudden approximation, the description of the initial and of the final state can be separated from each other.
\cite{fre09} Final-state multiple-scattering effects, dipole selection rules and, generally, all effects of the transition-matrix
elements as well as multiple scattering from the surface potential are fully included by describing the final state of the
photoelectron as a time-reversed low-energy-electron-diffraction (LEED) state. To simplify the solution of the Dyson-type
integral equation for the lesser Green's function, we adopt an ``atomic approximation'', i.e.\ we compute the atomic contribution
of the initial state only. This should be reasonable for the case of a pump-probe experiment where the time-dependent electronic
structure in a valence band is probed with an X-ray pulse addressing a core state. For this valence-pump---core-probe photoemission
\cite{piet08,Siff02,Hell12} a fully relativistic four-component formalism is necessary. Our approach should thus also make contact
with pump-probe photoemission from high-$Z$ materials and allows to study, e.g., dichroic effects in TR-PES.  

The paper  is organized as follows: The next section presents the general theory of time-resolved photoemission theory.
The relation to the conventional (equilibrium) one-step theory and other aspects are discussed in section 3. Section 4 is
devoted to the lesser Green's function which describes the time evolution of the system's initial state after a strong pump
pulse. Different types of pump-probe experiments are discussed in section 5, while the explicit formulation of time-resolved
photoemission within the one-step approach is worked out in section 6 for an X-ray probe addressing a core state. Section 7
provides a short summary.

\section{General theory of time-dependent photoemission spectroscopy}
We consider the electronic properties of a system specified by a Hamiltonian $\ca H$ and assume that the system's state
is a thermal state characterized by the inverse temperature $\beta$ and the chemical potential $\mu$ in the distant past
$t \to -\infty$. The grand canonical density operator is given by $\rho(-\infty) = Z^{-1} \exp(-\beta(\ca H - \mu N))$
where $N$ is the total particle number and $Z = \mbox{tr} \exp(-\beta(\ca H - \mu N))$ the partition function. 
Let $\ket{\Psi_{m}}$ be the eigenstates of $\ca H$ and $E_{m}$ the corresponding eigenenergies. We have 
\begin{eqnarray}
\rho(-\infty) = \sum_m p_m \ket{\Psi_m} \bra{\Psi_m}
\end{eqnarray}
with 
\begin{eqnarray}
p_m = \frac{1}{Z} e^{-\beta (E_m - \mu N_m)}  \: , 
\end{eqnarray}
where $N_{m}$ is the total particle number in the state $\ket{\Psi_{m}}$.

We consider a situation where the system is subjected to a strong light pulse, described by a light-matter interaction
Hamiltonian $\ca V(t)$, which drives the state $\rho(t)$ out of equilibrium. Typically, this pump pulse has a finite
duration and is followed by electronic relaxation processes on a femto-second time scale before slower relaxation
mechanisms involving lattice degrees of freedom become relevant. The time evolution of the mixed state
\begin{eqnarray}
\rho(t) = \sum_m p_m \ket{\Psi_m(t)} \bra{\Psi_m(t)}
\label{eq:state}
\end{eqnarray}
must be described non-perturbatively and is formally obtained from the time propagation of each state of the grand
ensemble, 
\begin{eqnarray}
\ket{\Psi_m(t)} = {\cal U}_{\rm tot}(t,-\infty) \ket{\Psi_m},
\end{eqnarray}
by means of the unitary time-evolution operator
\begin{eqnarray}
{\cal U}_{\rm tot}(t,t') = {\cal T} \exp\left( -i \int_{t'}^t {\cal H}_{\rm tot}(\tau) d\tau \right) \: ,
\end{eqnarray}
where ${\cal T}$ denotes chronological time ordering and where $\ca H_{\rm tot}(t) = \ca H + \ca V(t)$.

After some time delay $\Delta t$ following the pump, the non-equilibrium state is probed by a second pulse that is
described by an interaction term $\ca W(t)$. We assume that the probe pulse is non-zero for times $t>t_{0}$ and $t<t_{1}$. 
As we are merely interested in the electronic system properties, it is reasonable to express $\ca W(t)$ in terms of
electronic degrees of freedom only. Conceptually, the electronic structure can be subdivided into the states of primary
interest, namely occupied states and states in a certain energy window around the Fermi energy (or around $\mu$) on
the one hand and high-energy scattering states on the other. To address the former, we introduce $c_{\alpha}$ which
annihilates an electron in the one-particle basis state $\ket{\varphi_\alpha}$, while the latter are addressed by an
annihilator $a_{k}$ with a label $k$ for the one-particle scattering state that is occupied by the photoelectron. 
Therewith, 
\begin{eqnarray}
{\cal W}(t) = s_{\ca W}(t) \sum_{k, \alpha} (M_{k\alpha} a_k^\dagger c_\alpha + \mbox{H.c.}) \: , 
\label{eq:defw}
\end{eqnarray}
where $s_{\ca W}(t)$ describes the time profile of the probe pulse (and is non-zero for $t_{0} < t < t_{1}$ only) and
$M_{k,\alpha}$ are the transition-matrix elements for processes lifting a (low-energy) electron in the state
$\ket{\varphi_\alpha}$ to a high-energy state $k=(\ff k,\sigma)$ characterized by a wave vector $\ff k$ and a
spin projection $\sigma=\uparrow,\downarrow$. 

The probability $P_{k}(t)$ to detect a photoelectron with quantum numbers $k$ at time $t$ is given by the expectation
value, in the state $\rho(t)$, of the projector $\Pi(k)$ onto the subspace of all many-electron ``final'' states of
the form 
\begin{eqnarray}
\ket{f} = a_k^\dagger \ket{\Phi_n} \: .
\end{eqnarray}
Therewith we have adopted the sudden approximation and assumed that the Coulomb interaction of the (high-energy)
photoelectron with the low-energy part of the system can be neglected. $\ket{\Phi_n}$ is an arbitrary many-electron
state from an orthonormal basis set of the rest system (excluding the high-energy scattering states). We have 
\begin{eqnarray}
\Pi(k) = \sum \ket{f} \bra{f} = \sum_n a^\dagger_k \ket{\Phi_n} \bra{\Phi_n} a_k
\label{eq:pi}
\end{eqnarray}
and 
\begin{eqnarray}
P_k(t) 
= 
\ew{\Pi(k)}_{\rho(t)} = \tr \left( \rho(t) \Pi(k) \right) \: .
\end{eqnarray}
With Eqs.\ (\ref{eq:state}) and (\ref{eq:pi}), the time-dependent photoemission spectrum is obtained as
\begin{eqnarray}
P_k(t) 
= 
\sum_{m,n} p_{m} | \bra{\Phi_{n}} a_{k} \ket{\Psi_{m}(t)} |^{2} \: .
\label{eq:pes0}
\end{eqnarray}

The next task is to find the time dependence of the states $\ket{\Psi_{m}(t)}$ in the presence of the additional
probe pulse $\ca W(t)$, i.e.\ for times $t>t_{0}$. We have 
\begin{eqnarray}
\ket{\Psi_m(t)} = \ca U_{1}(t,-\infty) \ket{\Psi_m}
\label{eq:psim}
\end{eqnarray}
where
\begin{eqnarray}
{\cal U}_{1}(t,t') = {\cal T} \exp\left( -i \int_{t'}^t ({\cal H}_{\rm tot}(\tau) + \ca W(\tau)) d\tau \right) \: .
\end{eqnarray}
We assume that the probe pulse is sufficiently weak and treat the time evolution perturbatively.
To this end, let us introduce the corresponding S-matrix:
\begin{eqnarray}
{\cal S_W}(t,t_0) = \ca U_{\rm tot}(t_0,t) {\cal U}_{1}(t,t_0) \; .
\label{eq:smatrix}
\end{eqnarray}
Taking the time derivative, we immediately get
\begin{eqnarray}
i \frac{d}{dt} {\cal S_W}(t,t_0) = \ca U_{\rm tot}(t_0,t) {\cal W}(t) \ca U_{1}(t,t_0) \: .
\end{eqnarray}
Integration then yields
\begin{eqnarray}
{\cal S_W}(t,t_0) = 1-i \int_{t_0}^t dt' \ca U_{\rm tot}(t_0,t') {\cal W}(t') {\cal U}_{1}(t',t_0) \: .
\end{eqnarray}
Taking into account the term first order in $\ca W(t)$ only, we can replace ${\cal U}_{1}(t',t_0)$ by 
$\ca U_{\rm tot}(t',t_0)$ on the right-hand side. 
Therewith and using Eqs.\ (\ref{eq:psim}) and (\ref{eq:smatrix}), we find
\begin{widetext}
\begin{eqnarray}
\ket{\Psi_m(t)} 
\approx 
{\cal U}_{\rm tot}(t,-\infty) \Big( 1 -i \int_{-\infty}^t dt' {\cal U}_{\rm tot}(-\infty,t')
{\cal W}(t') {\cal U}_{\rm tot}(t',-\infty)\Big)
\ket{\Psi_m} \: .
\end{eqnarray}

Inserting this into Eq.\ (\ref{eq:pes0}) and recalling that $\ca W(t) = 0$ for $t<t_{0}$, yields
\begin{eqnarray}
P_k(t) 
= 
\sum_{m,n} p_{m} 
\left| \bra{\Phi_{n}} a_{k} 
\int_{t_{0}}^t dt' {\cal U}_{\rm tot}(t,t') {\cal W}(t') {\cal U}_{\rm tot}(t',-\infty) 
\ket{\Psi_m} 
\right|^{2} \: .
\label{eq:pes}
\end{eqnarray}
Here, we have also made use of the fact that the one-particle high-energy scattering states are, to a very
good approximation, unoccupied in $\ket{\Psi_m} = \ket{\Psi_m(-\infty)}$, i.e.\ that $a_k \ket{\Psi_m} \approx 0$.
Making once more use of the sudden approximation, we have 
\begin{eqnarray}
a_k  {\cal U}_{\rm tot}(t,t') = {\cal U}_{\rm tot}(t,t') a_k e^{-i \varepsilon(k)(t-t')} \: ,
\end{eqnarray}
where $\varepsilon(k)$ is the dispersion of the scattering state.
Furthermore,
\begin{eqnarray}
a_k {\cal W}(t') {\cal U}_{\rm tot}(t',-\infty) \ket{\Psi_m} = s_{\ca W}(t') \sum_{\gamma} M_{k \gamma}
c_\gamma {\cal U}_{\rm tot}(t',-\infty) \ket{\Psi_m} \; .
\end{eqnarray}
Therewith we arrive at
\begin{eqnarray}
P_k(t) = \sum_{m,n} p_m \left| \bra{\Phi_n} \int_{t_0}^t dt'
{\cal U}_{\rm tot}(t,t') e^{i \varepsilon(k) t'} s_{\ca W}(t') \sum_{\gamma} M_{k\gamma} c_\gamma 
{\cal U}_{\rm tot}(t',-\infty) \ket{\Psi_m} \right|^2 \: .
\end{eqnarray}
Reformulating this result by expanding the modulus square and using $\sum_{n} \ket{\Phi_n} \bra{\Phi_n} = \ff 1$,
we have:
\begin{equation}
P_k(t) 
=
\int_{t_{0}}^{t} \int_{t_0}^t dt'dt'' s_{\ca W}(t') s_{\ca W}(t'') e^{-i \varepsilon(k) (t'-t'')}
\sum_{\alpha\beta,m} M^\ast_{k\beta} M_{k\alpha} 
p_m \bra{\Psi_m} {\cal U}_{\rm tot}(-\infty,t') c^\dagger_\beta {\cal U}_{\rm tot}(t',t'')
c_\alpha {\cal U}_{\rm tot}(t'',-\infty) \ket{\Psi_m} \: .
\end{equation}
This can be written in a compact form by switching to the Heisenberg picture, i.e.
$c_\alpha(t) = {\cal U}_{\rm tot}(-\infty,t) c_\alpha {\cal U}_{\rm tot}(t,-\infty)$:
\begin{eqnarray}
P_k(t) 
=
\sum_{\alpha\beta} M^\ast_{k\beta} M_{k\alpha} 
\int_{t_0}^t dt' s_{\ca W}(t') \int_{t_0}^t dt'' s_{\ca W}(t'') e^{-i \varepsilon(k) (t'-t'')}
\langle c^\dagger_\beta(t') c_\alpha(t'') \rangle \: .
\label{eq:tdpes}
\end{eqnarray}
The one-particle correlation function is the lesser component of the Keldysh Green's function. It is given by
an equilibrium expectation value but is time inhomogeneous as the Heisenberg time dependence is governed by the
full and explicitly time-dependent Hamiltonian $\ca H_{\rm tot}(t) = \ca H + \ca V(t)$.
\end{widetext}

\section{Discussion}

Eq.\ (\ref{eq:tdpes}) was first derived by Freericks et al.\ in Ref.\ \onlinecite{fre09} in a similar way.
It nicely defines the main tasks of a theory of time-dependent pump-probe photoemission:
The problem consists in computing the lesser Green's function which describes the temporal evolution of the
electronic degrees of freedom after the pump pulse. As this is assumed to drive the system strongly out of
equilibrium, a linear-response approach or, generally, a perturbative calculation must be disregarded, and
ideally Dyson's equation with respect to $\ca V(t)$ should be solved without approximation. Within a picture of
effectively non-interacting electrons, this is equivalent with a time-dependent multiple-scattering approach
where all scattering events, at all times and in the entire lattice, are summed up. This is a formidable task. 

The second problem consists in the realistic computation of the transition-matrix elements in Eq.\ (\ref{eq:tdpes}).
For those we have to consider the light-matter interaction term $\ca W(t)$ in Eq.\ (\ref{eq:defw}). We apply the dipole
approximation which is well justified for sufficiently large wave lengths, i.e., for 
photon energies below about 10 keV. 
In the real-space representation and
using a relativistic four-component notation, needed for later purposes, we have
\begin{equation}
  W (\ff r, t) = W(t) = -s_{\ca W}(t) \mbox{\boldmath $\alpha$} \cdot {\bf A}_{0,\cal W} \: , 
\label{eq:wint}
\end{equation}
where ${\bf A}_{0,\cal W}$ denotes the spatially constant amplitude of the electromagnetic vector potential. 
The three components $\alpha_{k}$ of the vector $\mbox{\boldmath $\alpha$}$ are defined as the tensor product
$\alpha_{k} = \sigma_{1} \otimes \sigma_{k}$ for $k=1,2,3$ and where $\sigma_{k}$ denote the Pauli spin matrices.

While the lesser Green's function in Eq.\ (\ref{eq:tdpes}) describes time-dependent multiple-scattering from the pump
pulse ${\cal V}(t)$ in the ``initial'' state, the matrix elements also include the ``final'' state of the photoemission 
process. This is a one-particle scattering state that is characterized by $k=(\ff k,\sigma)$ and that has the
correct asymptotic behavior, i.e.\ is a simple plane wave ``at the detector'' far away from the system. 
Here, we will make use of the standard layer-KKR formalism \cite{Ebe10} to represent this 
state as a time-reversed LEED state. \cite{Pen74}

The real-space representation of the pump pulse ${\cal V}(t)$ has the same form as the probe. In both cases the excitation
is mediated by a light pulse where the corresponding electromagnetic field can essentially be described by a monochromatic
plane wave. 
The time profile of the pump pulse $s_{\ca V}(t)$ can be different from that of the probe $s_{\ca W}(t)$.
We assume that the dipole approximation is well justified for both, the pump and the probe.
The strength of
the two pulses can be largely different. This fact is encoded in the absolute magnitude of the vector potential. 

The conventional, i.e.\ equilibrium, expression of the photocurrent is easily re-derived from Eq.\ (\ref{eq:tdpes})
by assuming ${\cal V}(t) \equiv 0$. 
This immediately implies a time-homogeneous lesser Green's function. We also set
$s_{\ca W}(t) \equiv 1$ and consider the limits $t_0\to -\infty$ and $t\to \infty$. After a change of variables
$dt'dt''=dt_{\rm rel}dt_{\rm av}$ with $t_{\rm rel} = t'-t''$ and $t_{\rm av} = (t'+t'')/2$, one finds for the
transition probability per unit time:
\begin{eqnarray}
w_k 
& = &
\frac{P_k}{t-t_0}
=
\sum_{\alpha\beta} M^\ast_{k\beta} M_{k\alpha} 
\nonumber \\ &\times&
\int_{-\infty}^{\infty} dt_{\rm rel} \, e^{-i \varepsilon(k) t_{\rm rel}}
\langle c^\dagger_\beta(t_{\rm rel}) c_\alpha(0) \rangle \: .
\label{eq:fgr}
\end{eqnarray}
Using the spectral sum rule
\begin{eqnarray}
\ew{c^\dagger_\beta(t) c_\alpha(0)}
= 
\int_{\infty}^{\infty} d\omega \, f(\omega) A_{\alpha\beta}(\omega) e^{i\omega t} \; , 
\end{eqnarray}
where $f(\omega) = 1/(\exp(\beta \omega) +1)$ is the Fermi function, we have
\begin{eqnarray}
w_k = 2 \pi
\sum_{\alpha\beta} M^\ast_{k\beta} M_{k\alpha} 
f(\omega) A_{\alpha\beta}(\omega) \: ,
\end{eqnarray}
with $\omega = \varepsilon(k)$ and with the single-electron spectral density $A_{\alpha\beta}(\omega)$.
This is the well known golden-rule formula for the equilibrium photocurrent. \cite{Pen76}

\section{Initial-state Green's function}

The time-dependent correlation function in Eq.\ (\ref{eq:tdpes}) can be considered as a component of the one-particle
Keldysh Green's function $G_{\alpha\alpha'}(z,z')$. Some of its basic properties will be discussed here for the case
of effectively non-interacting electrons, i.e., we will not account for time-dependent correlations in the sense of
many-particle interactions. Generally, the Green's function is defined for arguments $z, z'$ on the Keldysh-Matsubara
\cite{keldys} contour $C$ in the complex time plane as
\begin{eqnarray}
i G_{\alpha\alpha'}(z,z') = \ew{{\cal T} c_\alpha(z) c^\dagger_{\alpha'}(z')} \: ,
\end{eqnarray}
where $\ca T$ denotes the contour ordering and where the expectation value refers to the ``free'' system $\ca H$
while the $z$ results from time evolution with the ``total'' Hamiltonian $\ca H_{\rm tot}(t) = \ca H + \ca V(t)$ with 
\begin{eqnarray}
{\cal H} =  \sum_{\alpha \alpha'} T_{0; \alpha \alpha'} c_\alpha^\dagger c_{\alpha'} 
\end{eqnarray}
and 
\begin{eqnarray}
{\cal V}(t) = \sum_{\alpha \alpha'} V_{\alpha \alpha'}(t) c_\alpha^\dagger c_{\alpha'} \: .
\end{eqnarray}
The Green's function can be obtained from Dyson's equation,
\begin{equation}
\ff G(z,z') = \ff G_0(z,z') + \int_C dz'' \ff G_0(z,z'') \ff V(z'') \ff G(z'',z') \: ,
\end{equation}
where fat symbols refer to matrices in the orbital indices $\alpha, \alpha'$. $G_{0; \alpha\alpha'}(z,z')$ denotes
the ``free'' Green's function for $\ca V(z) \equiv 0$ and is easily expressed in terms of the hopping matrix
$\ff T_{0}$ as \cite{bal11}
\begin{eqnarray}
&&i \ff G_0(z,z') = e^{-i \ff T_0 (z-z')}
\nonumber \\ 
&&\left[ \frac{\Theta_c(z,z')}{1+\exp(-\beta (\ff T_0 - \mu))} -
\frac{\Theta_c(z',z)}{\exp(\beta (\ff T_0 - \mu)) + 1} \right]  \: ,
\end{eqnarray}
where $\Theta_{c}(z,z')$ is the contour step function.
For our purposes it is sufficient to consider the lesser component of $\ff G(z,z')$ which is obtained for $z$ on
the upper and $z'$ on the lower branch of the Keldysh contour, i.e.\ $z$ is ``earlier'' than $z'$ on the contour
and thus $iG_{\alpha\alpha'}^{<}(t,t') = - \ew{c^\dagger_{\alpha'}(t') c_\alpha(t)}$. This can formally be written as
\begin{eqnarray}
i \ff G^<(t,t')
&=&
- {\cal T} e^{-i\int^t_{t_0}d\tau (\ff T_0 + \ff V(\tau))}
\frac{1}{e^{\beta(\ff T_0-\mu)}+1}
\nonumber \\ &\times&
{\tilde {\cal T}} e^{i\int^{t'}_{t_0}d\tau (\ff T_0+\ff V(\tau))} \: .
\label{eq:glesser}
\end{eqnarray}
where $\tilde {\cal T}$ is the anti-chronological time ordering.

We are seeking for a more suitable representation which also allows to set up a time-dependent multiple-scattering
approach. To this end we define the S-matrix ${\ff S_{\ca V}}$ related to (all-order) perturbation theory in the
pump pulse:
\begin{eqnarray}
{\ff S_{\ca V}}(t,t')
=
\ff U_0(t_0,t) \ff U(t,t') \ff U_0(t',t_0) \: .
\label{eq:defs}
\end{eqnarray}
Here $\ff U_{0}(t,t') = \exp(-i \ff T_{0} (t-t'))$ is the matrix representation of the free and
$\ff U(t,t') = {\cal T} \exp[ - i \int^t_{t'} d\tau (\ff T_0+\ff V(\tau)) ]$ the representation of the time-evolution
operator. The equation of motion for ${\ff S_{\ca V}}(t,t')$ is easily derived. We have
\begin{eqnarray}
i\frac{\partial}{\partial t} {\ff S_{\ca V}}(t,t')
=
\ff V_{t}(t) {\ff S_{\ca V}}(t,t') \: ,
\end{eqnarray}
where the double time-dependence appears due to the use of the interaction picture, 
$\ff V_t(t) = \ff U_0(t_0,t) \ff V(t) \ff U_0(t,t_0)$.
The corresponding integral equation reads as:
\begin{eqnarray}
{\ff S_{\ca V}}(t,t')
=
\ff 1 - i \int^t_{t'} dt'' \ff V_{t''}(t'')  {\ff S_{\ca V}}(t'',t') \: .
\end{eqnarray}
For $t>t_{0}$, the time-evolution matrices can also be expressed in terms of the retarded and advanced Green's functions,
$i \ff G^{\rm R}_0(t,t') \equiv i\Theta(t-t') \ff U_0(t,t')$ and $i \ff G^{\rm A}_{0}(t,t') = i \Theta(t' - t)\ff U_0(t,t')$, 
i.e.\
\begin{equation}
\ff V_{t}(t)
=
\ff G^{\rm A}_0(t_0,t)
\ff V(t)
\ff G^{\rm R}_0(t,t_0) \: .
\end{equation}
Therewith, we have (for $t,t' > t_{0}$)
\begin{equation}
{\ff S_{\ca V}}(t,t')
= 
\ff 1
-
i \int^t_{t'}d\tau \,
\ff G^{\rm A}_0(t_0,\tau)
\ff V(\tau) 
\ff G^{\rm R}_0(\tau,t_0) 
{\ff S_{\ca V}}(\tau,t') 
\: , 
\label{eq:ints}
\end{equation}
and with Eq.\ (\ref{eq:glesser}) and the definition of the S-matrix, we get
\begin{eqnarray}
i \ff G^<(t,t')
&=&
-
\ff G^{\rm R}_0(t,t_0)
\ff S_{\ca V}(t,t_0)
\frac{1}{e^{\beta(\ff T_0-\mu)}+1}
\nonumber \\ &\times&
\ff S_{\ca V}(t_0,t')
\ff G^{\rm A}_0(t_0,t') 
\: .
\end{eqnarray}
With Eq.\ (\ref{eq:glesser}) we can write 
\begin{equation}
i \ff G^<(t,t')
=
-
\ff G^{\rm R}(t,t_0)
\frac{1}{e^{\beta(\ff T_0-\mu)}+1}
\ff G^{\rm A}(t_0,t') 
\: .
\label{eq:ggg}
\end{equation}
The retarded Green function $\ff G^{\rm R}(t,t')$ can be obtained from the following integral equation 
\begin{equation}
\ff G^{\rm R}(t,t')
= 
\ff G^{\rm R}_0(t,t_0)
+
\int^t_{t'}d\tau \,
\ff G^{\rm R}_0(t,\tau)
\ff V(\tau) 
\ff G^{\rm R}(\tau,t') 
\: , 
\label{eq:dysongret}
\end{equation}
which derives from Eqs.\ (\ref{eq:defs}) and (\ref{eq:ints}). 
The advanced Green function is given by $\ff G^{\rm A}(t',t)=(\ff G^{\rm R}(t',t))^{\dagger}$.
This completes the formal calculation of the lesser Green's function.

\section{Discussion of different pump-probe spectroscopies}

The integral equation for $\ff G^{\rm R}(t,t')$ poses a time-dependent multiple-scattering problem for a three-dimensional
solid or, more realistically, for a semi-infinite system bounded by a surface. In combination with Eq.\ (\ref{eq:tdpes}),
it provides us with a quantitative description of different pump-probe experiments. Clearly, the solution appears as a
demanding task.

Typically the pump pulse $\ca V(t)$ excites electrons from occupied to unoccupied valence states below the vacuum level while
the probe pulse $\ca W(t)$, after some defined time delay, excites valence electrons from the explicitly time-dependent and
non-equilibrium state into high-energy scattering states such that they can escape into the vacuum. For those
valence-pump---valence-probe experiments, the numerical evaluation of the theory is most demanding. 

A simplification of the formalism is possible, however, for the important case of two-photon photoemission experiments (2PPE).
\cite{pick08,Schmidt10,Pick10,Wein10} Namely, from the theoretical perspective, a 2PPE experiment actually is just a pump-probe-type
experiment where the intensity of the pump pulse is comparable to the intensity of the probe. This means that Eq.\ (\ref{eq:dysongret})
can be treated perturbatively to a good approximation, and the series obtained by iteration can be cut by neglecting terms of the
order $\ca O(\ff V^{2})$, for example. This leaves us with an expression for the Green functions $\ff G^{\rm R}(t,t')$ that is
amenable to a straightforward numerical calculation since it is given in terms of $\ff G_0^{\rm R}(t,t') = \ff G_0^{\rm R}(t-t')$
only, i.e.\ in terms of the {\em equilibrium} retarded Green's function which is homogeneous in the time arguments. This quantity
is defined as the Fourier transformed of the following quantity, which is available from standard layer-dependent KKR
techniques: \cite{Macl89}
\begin{widetext}
\begin{eqnarray}
G^R_{0}({\ff r},{\ff r'},E) &=& -4ik\sum_{jn\Lambda} \Psi_{jn\Lambda}^{+}({\bf r_>})\Psi_{jn\Lambda}^{\dagger}({\bf r_<})
\nonumber \\  &-& 
\frac{4ik}{\Omega} \sum_{jn\Lambda \Lambda'} \Psi_{jn\Lambda}({\bf r})(t_{jn\Lambda})^{-1}
\Big( \int_{\Omega}d{\bf k} \tau^{nn}_{j\Lambda \Lambda'} - \delta_{\Lambda \Lambda'}t_{jn\Lambda'} \Big)
(t_{jn\Lambda'})^{-1} \Psi_{jn\Lambda'}^{\dagger}({\bf r'}) \: .
\nonumber \\
\label{eq:glkkr}
\end{eqnarray}
\end{widetext}
Here, we have switched to the real-space representation. $\Psi^+$ and $\Psi$ represent the single-site solutions for the
$n$-th cell in the $j$-th layer in a semi-infinite of slab geometry and $\tau$ is the KKR scattering path operator for the
$j$-th layer. $t$ denotes single-scattering matrix for the $n$-th cell in the $j$-th layer and $\Omega$ is the area of the
layer unit cell. Besides valence-pump---valence-probe and besides 2PPE experiments, a valence-pump---core-probe setup is
frequently used. \cite{piet08,Siff02,Hell12} Here the pump pulse excites valence-band electrons with a photon energy of a
few eV to unoccupied valence states. After a controlled time delay, the response of a core level is probed in a second step
with a corresponding X-ray probe pulse. The first part of this pump-probe experiment is described by $\ff G^{\rm R}(t,t')$
which must be obtained from the integral equation (\ref{eq:dysongret}) and accounts for the time evolution of the non-equilibrium
electronic structure after an intense pump pulse. Opposed to 2PPE, this time evolution usually cannot be captured in a
linear-response formalism, i.e., a perturbative approach expanding in $\ca V$ is not applicable here. There is, however, a
simplification suggesting itself for the case of a probe pulse addressing {\em core} electrons, namely to restrict oneself
to a single scattering center in the solution of the Eq.\ (\ref{eq:dysongret}) for the retarded Green functions
$\ff G^{\rm R}(t,t_0)$. For the case of an X-ray probe addressing a core state, the propagator refers to this core state only,
as is obvious from the central Eq.\ (\ref{eq:tdpes}). Within the one-step model, this results in an ``atomic contribution''
of the initial state to the full time-dependent photocurrent -- while a full summation over all multiple-scattering events is
included in the formalism for the final state. This shall be worked out in detail in the following section. In the case of
almost dispersion-free core states, this should be an excellent starting point. Clearly, the ultimate test case for the atomic
approximation will be the direct comparison with corresponding experimental data.\cite{piet08,Siff02,Hell05,Mel08,Hell12}

\section{One-step model of core-level pump-probe spectroscopy}

To formulate a one-step theory of valence-pump---core probe photoemission, a fully relativistic formalism for the final state
is necessary. We therefore rewrite Eq.\ (\ref{eq:tdpes}) in the real-space and a four-component spinor representation:
\begin{widetext}
\begin{equation}
P_k(t) 
= 
\int d^3 r' \int d^3 r'' \: f^{\dagger}_{k}({\ff r'}) W(t') 
\int_{t_0}^t dt' \int_{t_0}^t dt'' e^{-i \varepsilon(k) (t'-t'')}
G^<({\ff r'},t',{\ff r''},t'') W^{\dagger}(t'') f_{k}({\ff r''}) \: .
\end{equation}
Here, $G^<({\ff r'},t',{\ff r''},t'')$ is a $4\times 4$ Green's function matrix. $f_{k}$ represents a single-particle-like
final state of the photoelectron in form of a time-reversed LEED state and is a four-component spinor. The $4\times 4$ matrix
$W(t)$ is given by Eq.\ (\ref{eq:wint}). The lesser Green's function is obtained from Eq.\ (\ref{eq:ggg}) in real-space
representation,
\begin{equation}
G^<({\ff r},t,{\ff r'},t') = i \int dE~F(E) \int d^3 r''\int d^3 r'''
G^R(\ff r,t,\ff r'',t_0) \Psi_E({\ff r''})\Psi^{\dagger}_E({\ff r'''})
G^A(\ff r''',t_0,\ff r',t') 
\label{eq:gl}
\end{equation}
\end{widetext}
where $F(E) = 1 / (\exp(\beta (E - \mu)) +1)$ denotes the Fermi distribution function, and where the spinors $\Psi_E$ define
an orthonormal basis set with $T_0 \Psi_E = E  \Psi_E$. The occupied energy eigenstates $\Psi_{E}$ needed here can be obtained
from relativistic KKR theory. \cite{Ebe10} This also provides us with the retarded (and the advanced) Green's function
$G_{0}^{\rm R}$ (and $G_{0}^{\rm A}$), constructed as a tensor product of two four spinors. \cite{Bans99}

In the modern version of the KKR method, \cite{Ebe10} the electronic structure of a system, including valence as well as core
states, is directly and efficiently represented in terms of the retarded one-electron Green function. This appealing feature is
achieved by using multiple-scattering theory. The same multiple-scattering KKR technique is used in the context of photoemission
theory to construct the final state as a time-reversed LEED state. For these reasons, the KKR multiple-scattering formalism
provides the initial state Green's function and the final-state scattering state in a consistent way and on equal footing and
is thus the method of choice.

As argued in the preceding section, we will focus on the ``atomic'' contribution of the initial state to the total time-dependent
photoemission yield. For simplicity, we additionally assume a spherically symmetric single-cell potentials $v(\ff r) = v(r)$. 
Therewith, an ansatz separating radial and angular dependencies becomes convenient. For all types of Green's functions,
$G^<$, $G^{\rm R}$ and $G^{\rm A}$, we have 
\begin{equation}
G^{<,{\rm R,A}}({\ff r},t,{\ff r'},t')
=
\sum_{\Lambda \Lambda'}
g^{<,{\rm R,A}}_{\Lambda \Lambda'}(r,t,r',t') 
\chi_{\Lambda}({\bf \hat r}) \chi^{\dagger}_{\Lambda'}({\bf \hat r'}) 
\label{eq:radialan}
\end{equation}
where $\chi_{\Lambda}({\bf \hat r})$ denote the relativistic spin-angular functions \cite{Ros61} with the spin-orbit ($\kappa$)
and the magnetic ($\mu$) quantum numbers \cite{Ros61} combined to $\Lambda$=($\kappa, \mu$). Using this, the radial parts of the
quantities in Eq.\ (\ref{eq:gl}) are related by
\begin{widetext}
\begin{equation}
g^<_{\Lambda \Lambda'}(r,t,r',t')
=
i \int dE~F(E) \sum_{\Lambda'',\Lambda'''} \int dr'' r''^2 \int dr''' r'''^2~
g^{\rm R}_{\Lambda \Lambda''}(r,t,r'',t_0) 
\Psi_{E,\Lambda''}(r'')
\Psi^{\dagger}_{E,\Lambda'''}(r''') g^{\rm A}_{\Lambda''' \Lambda'}(r'',t_0,r',t') \: .
\end{equation}
This equation is considerably simpler and well amenable to a numerical approach based on the KKR formalism.

Likewise the central Dyson equation for the retarded Green's function, Eq.\ (\ref{eq:dysongret}), can be simplified. 
In the real-space representation we have
\begin{equation}
G^{{\rm R}}({\ff r},t,{\ff r'},t')
= 
G_0^{{\rm R}}({\ff r},t,{\ff r'},t') 
- 
\int_{t'}^{t}dt'' 
s_{\ca V}(t'')
\int d^3r'' G_0^{{\rm R}}({\ff r},t,{\ff r''},t'') \mbox{\boldmath $\alpha$} \cdot {\bf A}_{0,\cal V} 
G^{{\rm R}}({\ff r''},t'',{\ff r'},t') \: .
\end{equation}
Here, as for the probe, we assume that the dipole approximation is valid:
\begin{eqnarray}
V({\ff r},t)
=
-
s_{\ca V}(t) 
\mbox{\boldmath $\alpha$} \cdot {\bf A}_{0,\cal V} \: .
\label{eq:dipole}
\end{eqnarray}
$\ff A_{0, \ca V}$ denotes the constant amplitude of the vector potential of the pump field, and $s_{\ca V}(t)$ is the time
profile of the pump pulse. Using the ansatz Eq.\ (\ref{eq:radialan}) to get the atomic contribution, we find a coupled system
of Volterra integral equations of the second kind for the radial part of the retarded Green's function: 
\begin{eqnarray}
g^{\rm R}_{\Lambda \Lambda'}(r,t,r',t') 
=
g^{\rm R}_{0,\Lambda \Lambda'}(r,t,r',t')
-
\sum_{\Lambda''} \int_{t'}^t dt''~s_{\ca V}(t'') \int dr'' r''^2~
g^{\rm R}_{0,\Lambda \Lambda''}(r,t,r'',t'') \frac{dv(r'')}{dr''} g^{\rm R}_{\Lambda'' \Lambda'}(r'',t'',r',t')~. 
\nonumber \\  
\label{eq:gret}
\end{eqnarray}
We also made use of a representation of the dipole operator, see Eq.\ (\ref{eq:dipole}), in terms of the gradient of the
cell potential $v(r)$. Again, the resulting system of equations is considerably simplified and can be implemented numerically.
At this point the calculation of the atomic contribution is complete. 

We finally combine the results for the initial state with the layer-KKR multiple-scattering approach describing the final state. 
This provides us with the time-dependent photoemission yield in the following form:
\begin{eqnarray}
P_{k}(t)
=
\sum_{\Lambda \Lambda' \Lambda'' \Lambda''' jn}~A^{\dagger}_{jn\Lambda}~
D_{\Lambda \Lambda'}~M_{\Lambda \Lambda' \Lambda'' \Lambda'''}(t)~D_{\Lambda'' \Lambda'''}^{\dagger}~A_{jn\Lambda'''}~.
\label{eq:photo}
\end{eqnarray}
Here, the radial matrix elements are defined as 
\begin{eqnarray}
M_{\Lambda \Lambda' \Lambda'' \Lambda'''}(t) &=& \int_{t_0}^tdt' \int_{t_0}^tdt''~s_{\ca W}(t')s_{\ca W}(t'')
e^{-i\varepsilon(k) (t'-t'')} \int dr'r'^2 \int dr'' r''^2~
\nonumber \\ &&
\phi_{\Lambda}^{f\dagger}(r')~\frac{dv(r')}{dr'}~g^<_{\Lambda' \Lambda''}(r',t',r'',t'')~
\frac{dv(r'')}{dr''}~\phi_{\Lambda'''}^f(r'')~,
\label{eq:radial}
\end{eqnarray}
where $v$ is the spherical single-cell potential. Furthermore, $A$ denote the spherical coefficients of the high-energy wave field:
\begin{eqnarray}
A_{jn\Lambda}=\sum_{\Lambda'} A^{(o)}_{jn\Lambda'} ~(1-X)^{-1}_{\Lambda \Lambda' n}~.
\label{eq:fin1}
\end{eqnarray}
The effect of multiple-scattering within the $j$-th layer can be represented by a matrix X, and the bare coefficients $A^{(o)}$
are given by
\begin{eqnarray}
A^{(o)}_{jn\Lambda'}= \sum_{{\bf g}s} 4\pi i^{l'}(-2s)(-)^{\mu' -s} C^{\kappa' \mu'}_{s} \big[ u^{+}_{j{\bf g}s}
Y^{s-\mu'}_{l'} (\widehat{k^{+}_{2{\bf g}}}) e^{i{\bf k}^{+}_{2{\bf g}} \cdot {\bf r}_{n}}+
u^{-}_{j{\bf g}s} Y^{s-\mu'}_{l'} (\widehat{k^{-}_{2{\bf g}}})e^{i{\bf k}^{-}_{2{\bf g}} \cdot {\bf r}_{n}} \big]~.
\label{eq:fin2}
\end{eqnarray}
\end{widetext}
${\ff r}_{n}$ is the distance-vector from the origin to the position of the $n$-th atom in the layer-unit cell.
The plane-wave amplitudes $u^{+}$ and $u^{-}$ can be calculated recursively by standard (KKR) multiple scattering
techniques. \cite{KKR,Bra96} $D$ are the relativistic angular dipole matrix elements: \cite{Bra96}
\begin{eqnarray}
D_{\kappa \mu \kappa' \mu'}= \sum_{s=\pm \frac{1}{2}} (-2s)~
C^{\kappa \mu}_{s}~D^{\rm NR}_{l,\mu -s,l',\mu' +s}~C^{\kappa' \mu'}_{-s} \: ,
\end{eqnarray}
where $C^{\kappa \mu}_{s}$ denote the Clebsch-Gordan coefficients and the $D^{\rm NR}_{l,\mu -s,l',\mu' +s}$ represent the
non-relativistic angular matrix elements \cite{Pen76} which are given by 
\begin{equation}
  D^{\rm NR}_{LL'} = \frac{4\pi}{3} A_0 \: Y_{1m}^\ast(\hat{A_0}) \:
  C_{lml'm'1(-m-m')} 
\end{equation}
in terms of the Gaunt coefficients $C$. 

In Eq.\ (\ref{eq:photo}), the $k$ dependence, i.e., the dependence on the quantum numbers of the photoelectrons enters
$P_{k}(t)$ through the wave field describing the final state, Eqs.\ (\ref{eq:fin1}) and (\ref{eq:fin2}) but also through
the time-dependent phase factor via $\varepsilon(k)$ absorbed in the radial matrix elements, Eq.\ (\ref{eq:radial}).
The main experimental control parameter, the time delay, enters the theory through the time distance between the two pulses
with profiles $s_{\ca V}(t)$ (pump pulse) and $s_{\ca W}(t)$ (probe pulse) where the former appears in the integral equation
(\ref{eq:gret}) for the initial-state retarded Green's function while the latter appears more explicitly in Eq.\ (\ref{eq:radial}).

\section{Summary}

Concluding, we have presented a theoretical frame for time-resolved pump-probe photoemission involving electronic degrees of
freedom only. The theory addresses systems where explicit Coulomb correlations can be neglected safely and which can be treated
within a picture of effectively independent electrons moving in a one-particle potential that is obtained from standard
band-structure calculations. The present approach aims at an {\em ab initio} description of photoemission from real materials
that may complement pure model studies for strongly correlated systems carried out previously. 

The description of the transition induced by the probe pulse as well as the final high-energy scattering state of the
photoelectrons has been done in a fully relativistic, four-component formalism which resolves all quantum numbers of the
photoelectron. Thereby, the theory covers ultraviolet as well as soft or hard X-ray photon energies and spans the same regime
as the conventional equilibrium theory of angle-resolved photoemission. The central idea of our paper is to straightforwardly
extend the traditional and highly successful one-step formulation. Formally, this gives access to time-resolved intensity
distributions, magnetic linear and circular dichroism from real systems like simple metals, but also from complex ordered
compounds. While the formal construction of the theory has been achieved, our long-term goal is to implement a corresponding
efficient numerical tool that makes direct contact to experimental data.

In the equilibrium theory one may distinguish between different contributions to the total photocurrent, i.e., the atomic
contribution, where the propagation of the remaining hole is neglected, and different (intra- and inter-layer) multiple-scattering
corrections to the atomic contribution. In the case of X-ray photoemission from core states, the atomic contribution represents
the overwhelming part to the photocurrent -- the dispersion of the core states can be neglected in most cases. For the
non-equilibrium theory, however, the same ``atomic'' or ``single-cell'' approximation is more questionable -- even if an X-ray
probe pulse exciting core electrons is considered. The reason is that, typically, the main interest is on the non-equilibrium
dynamics of an excited state of the valence electrons. This is induced by a low-energy but strong pump pulse exciting electrons
from occupied to unoccupied states within the valence band. While the motion of the core hole can safely be restricted to a
single cell, there is a non-trivial valence-electron dynamics spreading over the lattice. A part of that dynamics is expected to
be affected by inter-atomic multiple-scattering at the pump pulse. Here, the atomic approximation has been adopted for reasons
of simplicity. Whether or not this yields the dominant contribution as compared to the inter-atomic multiple-scattering, is an
open question that must be addressed in the future.
 
\begin{acknowledgments}  
Financial support by the Deutsche Forschungsgemeinschaft through the Sonderforschungsbereich 925 
(project B5) and through projects P1 and P3 within FOR 1346, Eb-154/23, Eb-154/26 is gratefully acknowledged.
\end{acknowledgments}

\end{document}